\documentclass[10pt,conference]{IEEEtran}
\usepackage{dannychen}

\newcommand{\bitstring}[2]{\hat b_{#1} \dots \hat b_{#2}}

\begin{document}
\title{Approximate Quantum Circuit Reconstruction}

\author{\IEEEauthorblockN{Daniel Chen\IEEEauthorrefmark{1}, Betis Baheri\IEEEauthorrefmark{2}, Vipin Chaudhary\IEEEauthorrefmark{1}, Qiang Guan\IEEEauthorrefmark{2}, Ning Xie\IEEEauthorrefmark{3}, Shuai Xu\IEEEauthorrefmark{1}}
\IEEEauthorblockA{\IEEEauthorrefmark{1} Department of Computer and Data Science \\ Case Western Reserve University, Cleveland, Ohio 33549\\
Email: \{txc461, vxc204, sxx214\}@case.edu}
\IEEEauthorblockA{\IEEEauthorrefmark{2} Department of Computer Science\\ Kent State University, Kent, Ohio 44240\\ Email: \{bbaheri, qguan\}@kent.edu}
\IEEEauthorblockA{\IEEEauthorrefmark{3} Knight Foundation School of Computing and Information Sciences \\ Florida International University, Miami, FL 33199 \\ Email: nxie@cis.fiu.edu}
}

\maketitle

\begin{abstract}
    Current and imminent quantum hardware lacks reliability and applicability due to noise and limited qubit counts. Quantum circuit cutting --- a technique dividing large quantum circuits into smaller subcircuits with sizes appropriate for the limited quantum resource at hand --- is used to mitigate these problems. However, classical postprocessing involved in circuit cutting generally grows exponentially with the number of cuts and quantum counts. This article introduces the notion of approximate circuit reconstruction. Using a sampling-based method like Markov Chain Monte Carlo (MCMC), we probabilistically select bit strings of high probability upon reconstruction. This avoids excessive calculations when reconstructing the full probability distribution. Our results show that such a sampling-based postprocessing method holds great potential for fast and reliable circuit reconstruction in the NISQ era and beyond. 
\end{abstract}

\section{Introduction}

Quantum computers are believed to enable polynomial or even exponential speed-up over classical computers for many classes of problems \cite{bernstein1997quantum}. However, current quantum computing hardware, also known as NISQ computers, is characterized by its lack of scalability and reliability \cite{preskill2018quantum}. To effectively handle noise, the quantum algorithm community has leveraged classical resources to aid computation, which led to the emergence of quantum-classical hybrid algorithms~\cite{bharti2022noisy}. In particular, variational quantum circuits has so far received great attention \cite{cerezo2021variational}, with well-known examples like quantum approximate optimization algorithm (QAOA) \cite{farhi2014quantum} and variational quantum eigensolver (VQE) \cite{peruzzo2014variational} \cite{kandala2017hardware}. Such variational circuits perform expensive calculations (such as solving NP classical optimization or simulating Hamiltonians) using a quantum computer and update the circuit parameters using classical optimizers like COBYLA \cite{powell1994direct} or gradient descent \cite{schuld2019evaluating}. 

Hybrid algorithms mentioned above have enjoyed some success in the NISQ era, showing their robustness against noisy computation \cite{xue2021effects}. At the same time, escaping the issue of vanishing gradients, or barren plateaus, induced by noise, remains a problem \cite{wang2021noise}. Also, the size of NISQ computers poses barriers for most real applications. Thus, Peng \textit{et~al.} \cite{peng2020simulating} developed a method for dividing any quantum circuit into multiple smaller subcircuits that can be executed independently. Then, one could use the information gathered from each subcircuit to reproduce the theoretical outcome from running the full, uncut circuit. In addition to making evaluation of large circuits possible, it is also shown that evaluating smaller circuits improves fidelity while being more robust to noise \cite{tang2021cutqc} \cite{ayral2020quantum} \cite{perlin2021quantum}. 

Despite the theoretical success, reconstructing the probability distribution of the full circuit is time-consuming. More specifically, the classical post-processing procedures generally scale exponentially with respect to the size of the circuits. If one desires the classical probability distribution of the full circuit, there will be an additional exponential factor with respect to the circuit size. One way to avoid this scaling is to design algorithms that reduce the number of cuts needed \cite{saleem2021quantum}. However, we take a more direct approach in attempting to lower the reconstruction time complexity. Similar to how probabilistic methods are used to provide approximate solutions to problems that are difficult to solve exactly, we propose the concept of an \textit{approximate circuit reconstruction}. Instead of exploring exponentially growing state-space via brute force, we probabilistically explore the space of bit-strings and sample the points that are of higher likelihood. This procedure is asymptotically correct as the number of samples taken increases. We also found a significant reduction in run time, shedding light on the potential utility of applying this technique for larger, more practical problems. 

The contribution of this work can be summarized as follows:
\begin{itemize}
    \item We introduce the notion of \textit{approximate circuit reconstruction} using Monte Carlo methods for reconstructing measurement outcomes of subcircuits after cutting.
    \item Our model puts no explicit restriction on the run time: one can decide the trade-off between speed and accuracy depending on one's resources.
    \item The naive implementation was able to reconstruct the circuit outcome at speed an order of magnitude faster without losing much accuracy, motivating further development and optimization of this method.
\end{itemize}

\section{Quantum Circuit Cutting}\label{sec:circcut}

\begin{figure*}[!t]
\centering
\subfloat[Circuit prior to cutting. The cut location is specified at the red cross. Upon performing the cut, the circuit naturally separates into two independent subcircuits that can be run in parallel.] {\includegraphics[width=2.5in]{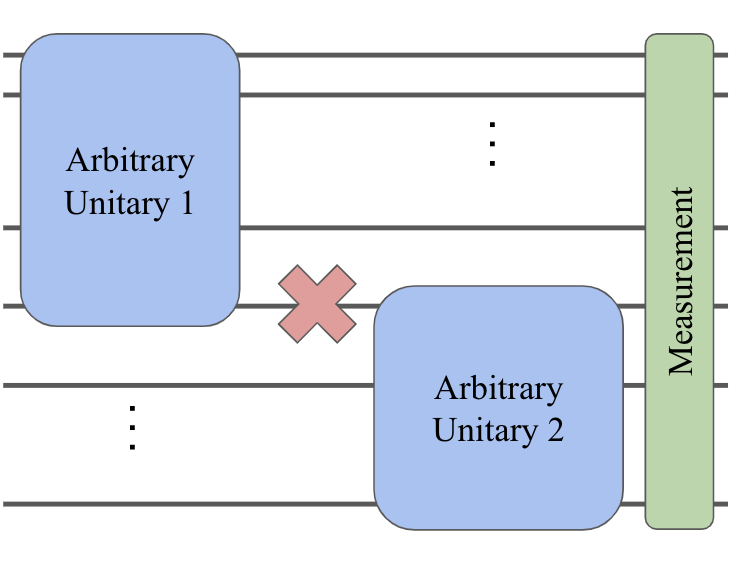}
\label{fig_first_case}}
\hspace{2cm}
\subfloat[Resulting subcircuits. (top) The last qubit of the upstream subcircuit, denoted by $\hat a$, requires measurements in basis $\beta$. The first qubit of downstream subcircuit (bottom) requires preparing the eigenstate of $\beta$ corresponding to eigenvalue $e$.]{\includegraphics[width=2.5in]{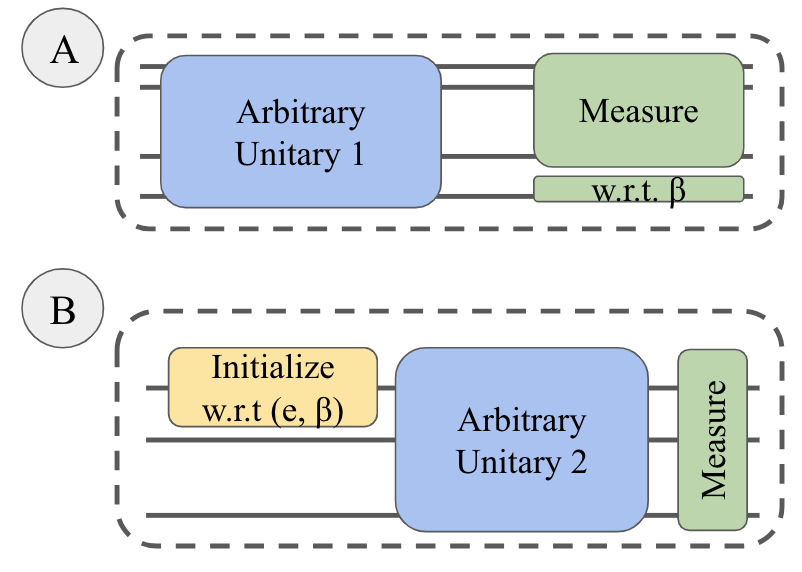}
\label{fig_second_case}}
\caption{A pictorial representation of the circuit cutting procedure.}
\label{fig_sim}
\end{figure*}

In short, the idea of quantum circuit cutting is separating circuits by applying a set of measurements and state preparation such that, when combined, results in a trivial action. Consider the Pauli matrices.
\begin{align}
    \sigma_x = 
    \begin{pmatrix}
    0 & 1 \\ -1 & 0 
    \end{pmatrix}, ~~ \sigma_y = 
    \begin{pmatrix}
    0 & -i \\ i & 0 
    \end{pmatrix}, ~~ \sigma_z = 
    \begin{pmatrix}
    1 & 0 \\ 0 & -1
    \end{pmatrix}
\end{align}
Together with $I=\begin{pmatrix} 1 & 0 \\ 0 & 1 \end{pmatrix}$, these four matrices form an orthonormal basis over the set of $2\times2$ complex matrices. We can apply a combination of measurements and initializations in different Pauli bases such that the state itself is unchanged. For the formalism presented in this paper, readers should refer to \cite{ayral2020quantum}.

We will begin by considering the case that there is one cut. Suppose there is an $m$-qubit quantum circuit $C$, which can be cut into subcircuits $A, B$ (c.f. Figure \ref{fig_sim}). Subcircuit $A$ has $n$ qubits whose measurement outcome in the computational basis will be represented as $\bitstring 1 {n-1} \hat a$. We can compactly store the set of outcomes of subcircuit $A$ with a rank-3 tensor $p_A\ind{\bitstring 1 {n-1}, \hat a, \beta}$. The first index denotes the measurement outcome of the first $n-1$ qubits. The second index refers to the extra ``connection'' bit $\hat a \in \{0,1\}$. There is always this extra bit upstream of the cut that will be measured in different bases. It naturally exists as part of the subcircuit and does not directly correspond to the circuit output. The third index refers to the measurement basis, $\beta \in \mathcal B = \{\sigma_x, \sigma_y, \sigma_z\}$.

Similarly, subcircuit $B$ is a $m-n+1$ qubit circuit with measurement outcomes $\bitstring n m$. This can again be written compactly as a rank-3 tensor $p_B\ind{\bitstring n m, e, \beta}$. The first index represents the measurement outcome of subcircuit $B$. The second and third indices are related: $e$ denotes the eigenbasis, with respect to $\beta \in \mathcal B$, that the qubit immediately downstream of the cut is initialized to. For example, if $e = 0$ and $\beta = \sigma_x$, then the qubit is initialized to the $\ket +$ state (the eigenvector of $\beta$ corresponding to eigenvalue $1$).

The probability distribution $P$ corresponding to the outcome can simply be written as a vector of $2^{m}$ bins, each corresponding to a measurement outcome. Finding the probability of an outcome $\hat b_1 \hat b_2 \dots \hat b_m$ from its subcircuits is represented in the following equation, which one can think of as a tensor contraction over indices $\hat a, e, \beta$.
\begin{align}\label{eq:recon}
    &P\ind{\bitstring 1 {n-1}, \bitstring n m} \nonumber \\
    &~~~~ = \sum_{\hat a,e,\beta} \gamma \ind{\hat a,e,\beta}p_A\ind{\bitstring 1 {n-1}, \hat a, \beta} p_B\ind{\bitstring n m, e, \beta}
\end{align}
and the rank-3 tensor $\gamma$ is defined as follows.
\begin{align}
    \gamma\ind{\hat a,e,\beta} = \begin{cases} 2\delta_{\hat a,e} - 1 &\text{if $\beta = \sigma_x$ or $\sigma_y$} \\ 2\delta_{\hat a,e} &\IF \beta = \sigma_z \end{cases}
\end{align}

The same formalism can also be extended to the case of multiple cuts with additional bookkeeping. Let there be $K$ cuts and $K+1$ subcircuits. For each cut, there will be an associated triple $(\hat a_k, e_k, \beta_k)$ representing the connection bit, the eigenvalue, and measurement/initialization basis respectively. Furthermore, each subcircuit tensor will have its respective elements, namely, $\hat a_k$ or $e_k$, depending on whether it is the upstream or downstream circuit. The matrix-product state formalism gives a concise representation for generalized circuit cutting (cf. \cite{ayral2020quantum, schollwock2011density, orus2014practical}). 

This contraction involves summing over 12 terms --- there are two possibilities each for $\hat a$ and $e$, three for $\beta$ --- which means that simulating the probability of obtaining one particular bit string takes only constant time. However, the naive method for constructing the output of the full distribution requires querying the probability of each bit string, which there are exponentially many, one by one. Also, since the representation above is only exact given completely accurate state tomography, the realistic recovered ``probability distribution'' might contain negative probabilities or are not normalized due to finite-shot noise and hardware noise (if ran on a physical device). Moreover, the time complexity of reconstructing the full distribution also grows exponentially with respect to the number of cuts as three additional indices are to be contracted for each additional cut. These unfavorable scaling factors inhibit circuit cutting from being used at its full potential. 

\section{Approximate Circuit Reconstruction}\label{sec:approx}

This section introduces the idea of an \textit{approximate circuit reconstruction} that seeks to avoid exponential scaling. For ease of exposition, we will demonstrate the reconstruction for the case of a single cut and two cuts; generalizing to arbitrary cuts is straightforward but tedious. 

The high-level idea is as follows: our goal is to build an approximate distribution with polynomially many queries from an exponentially growing state space. Since the weight of each bit string is unknown a priori, there is not a good deterministic strategy for approximating the distribution. Thus, we introduce a naive version of the Metropolis-Hastings (MH) algorithm, which is a type of Markov Chain Monte Carlo (MCMC) method. Let $f(\hat x)$ be a probability distribution, $\hat x \in \{0,1\}^n$, that we would like to sample from. Moreover, the unnormalized likelihood ratio between adjacent points in the support of the distribution can be computed efficiently. To avoid computing $f(\hat x)$ on every $x$ in the support, we perform random walks on a Markov chain defined on $ \{0,1\}^n$ with transition kernel $g(\hat y | \hat x)$. Then, at each step of the random walk, we move to state $\hat y$ from state $\hat x$ probabilistically with the probability of transition increasing if the ratio of likelihoods increases. For details of MH algorithms and Markov chain mixing, readers are referred to~\cite{robert1999metropolis, gelman1995bayesian, brooks2011handbook,levin2017markov}.

\subsection{Example: one-cut case}

Consider the case that a single cut has been made, producing outcomes $p_A$ and $p_B$ from the respective subcircuits where equation \eqref{eq:recon} describes the relationship between the subcircuits and the full circuit. 
In our case, each state $\hat x$ represents a possible output from the full quantum circuit and the distribution we would like to sample from is the one described in equation \eqref{eq:recon}. The complete procedure is described in Algorithm \ref{alg:onecut}.

\begin{algorithm}
\caption{Metropolis-Hastings algorithm for the one-cut case}\label{alg:onecut}
\KwIn{N: number of samples, BI: burn-in}
\KwOut{S: histogram of accepted bit strings}
$S \gets \{\}, ~~ n \gets 0$ \\
$\hat x \gets$ random bit string of length $m$ \\
\While {$n < N$} {
    $\hat y \gets$ bit string of length $m$ that is different with $x$ in only $1$ entry \\
    Let $r = \min\{1, P\ind {\hat y} / P \ind {\hat x}\}$ (cf. eq. \eqref{eq:recon}) \\
    \If {$u \sim \text{Uniform(0,1)} < r$} {
        $\hat x \gets \hat y$ \\
    }
    \If {$n > \text{BI}\cdot N$} {$S[\hat x] \gets S[\hat x] + 1$ }
        $n \gets n + 1$
}
\Return S
\end{algorithm}

Here, the transition kernel is one that flips one randomly-chosen bit from the conditioned bit string.
This procedure is easily parallelizable in the case of one cut. One can establish multiple \textit{chains}: implementation of the same algorithm with different initial conditions running in parallel. This comes at the advantage of not only time complexity, but also prevents improper chain mixing and guarantees convergence in the sample size limit.

\subsection{Example: two-cut case}

\begin{figure}
    \centering
    \includegraphics[width=.9\linewidth]{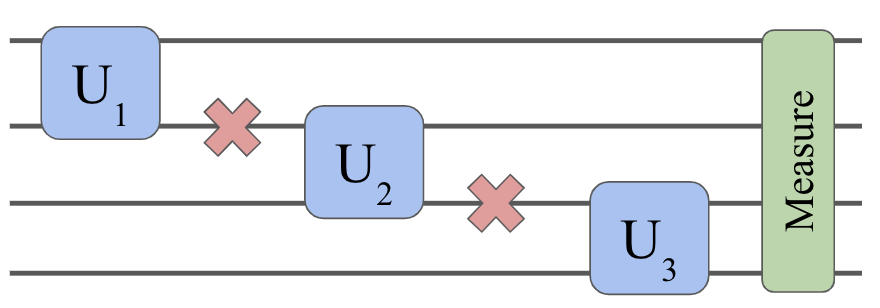}
    \caption{Example of a circuit with two cuts. Subcircuits $A$, $B$, and $C$ can be arbitrary.}
    \label{fig:twocut}
\end{figure}

Now consider the case where two cuts are made, resulting in three subcircuits $A$, $B$, and $C$ of size $n_1, n_2, n_3$ respectively. For concreteness, suppose the cuts follow the cascade-like structure depicted in Figure \ref{fig:twocut}. We will refer to the cut between subcircuit $A$ and $B$ as the first cut, and the one between $B$ and $C$ as the second. Subcircuit $A$ is upstream of $B$, meaning that it will need additional measurements at the location of the cut, denoted by indices $\hat a_1$ and $\beta_1$. Subcircuit $C$ is downstream of $B$, so it will need additional initializations at the location of the cut, represented by the indices $e_2, \beta_2$. Subcircuit $B$ is downstream of $A$ and upstream of $C$, which means that it will contain information about both subcircuits, meaning that $p_B$ will be a rank-5 tensor with indices $e_1, \beta_1, \hat a_2, \beta_2$ along with the measurement bit strings. 

We begin to realize the exponential scaling starting in the two-cut case. The exponential growth comes from the fact that, for each cut, we produce a pair of measurement-initialization combinations. Quantum information must be passed through each cut, which means that we must sum over all measurement-initialization pairs for every cut, resulting in roughly $2^{\text{poly}K}$ in runtime complexity. Here, we will propose two types of approximate reconstruction. The first obeys the exponential scaling with $K$ to gain accuracy at the cost of efficiency. It involves running six chains, one for each index pair $(e_1, \beta_1)$, as shown in Algorithm \ref{fig:twocut}.

\begin{algorithm}
\caption{Metropolis-Hastings algorithm for the two-cut case}\label{alg:twocut}
\KwIn{N: number of samples, BI: burn-in}
\KwOut{S: histogram of accepted bit strings}
$p_{BC} \gets \texttt{zeros}\ind{\hat b_1 \dots \hat b_{n_2 + n_3 -1}, e_1, \beta_1} $\\
$n \gets [0, 0, 0, 0, 0, 0]$ \\
$[\hat x_1, \dots, \hat x_6] \gets$ random bit strings of length $n_2 + n_3 - 1$ \\
\While {$\min(n) < N$} {
    \For {$(e_1, \beta_1) \in \{0,1\}\times \mathcal B$} {
        $\hat y \gets$ bit string of length $n_2 + n_3 - 1$ that is different with $\hat x$ in only $1$ entry \\
        Let $r = \min\{1, P\ind {\hat y} / P \ind {\hat x}\}$ \\
        \If {$u \sim \text{Uniform(0,1)} < r$} {
            $\hat x_i \gets y$ \\
        }
        \If {$n > \text{BI}\cdot N$} {$p_{BC}\ind{\hat x, e_1, \beta_1} \gets p_{BC}\ind{\hat x, e_1, \beta_1} + 1$ }
        $n[i] \gets n[i] + 1$
    }
}
\Return \texttt{reconstructOneCut}$(p_A, p_{BC})$
\end{algorithm}
The function \texttt{reconstructOneCut} calls Algorithm \ref{alg:onecut}. Here, we simply avoided the exponential scaling in the space of bit strings. However, the scaling with respect to the number of cuts will be preserved as more indices are added into the middle for-loop. As an attempt to avoid this scaling as well, we propose to select which index to contract uniformly at random. The algorithm is presented in Algorithm \ref{alg:twocutrand}

\begin{algorithm}
\caption{Metropolis-Hastings algorithm for the two-cut case with randomized indexing}\label{alg:twocutrand}
\KwIn{N: number of samples, BI: burn-in}
\KwOut{S: histogram of accepted bit strings}
$p_{BC} \gets \texttt{zeros}\ind{\hat b_1 \dots \hat b_{n_2 + n_3 -1}, e_1, \beta_1} $\\
$n \gets [0, 0, 0, 0, 0, 0]$ \\
$[\hat x_1, \dots, \hat x_6] \gets$ random bit strings of length $n_2 + n_3 - 1$ \\
\While {$\min(n) < N$} {
    $e_1 \gets 0,1$ uniformly at random \\
    $\beta \gets \sigma_x, \sigma_y, \sigma_z$ uniformly at random\\
    $\hat y \gets$ bit string of length $n_2 + n_3 - 1$ that is different with $\hat x$ in only $1$ entry \\
    Let $r = \min\{1, P\ind {\hat y} / P \ind {\hat x}\}$ \\
    \If {$u \sim \text{Uniform(0,1)} < r$} {
        $\hat x_i \gets y$ \\
    }
    \If {$n > \text{BI}\cdot N$} {$p_{BC}\ind{\hat x, e_1, \beta_1} \gets p_{BC}\ind{\hat x, e_1, \beta_1} + 1$ }
    $n[i] \gets n[i] + 1$
}
\Return \texttt{reconstructOneCut}$(p_A, p_{BC})$
\end{algorithm}

The asymptotic property is preserved even if we randomly choose which index to contract over. However, the rate at which it converges is slowed, which is the usual trade-off between efficiency and accuracy.

\subsection{Generalizing to an arbitrary number of cuts}

\begin{figure}
    \centering
    \includegraphics[width=.7\linewidth]{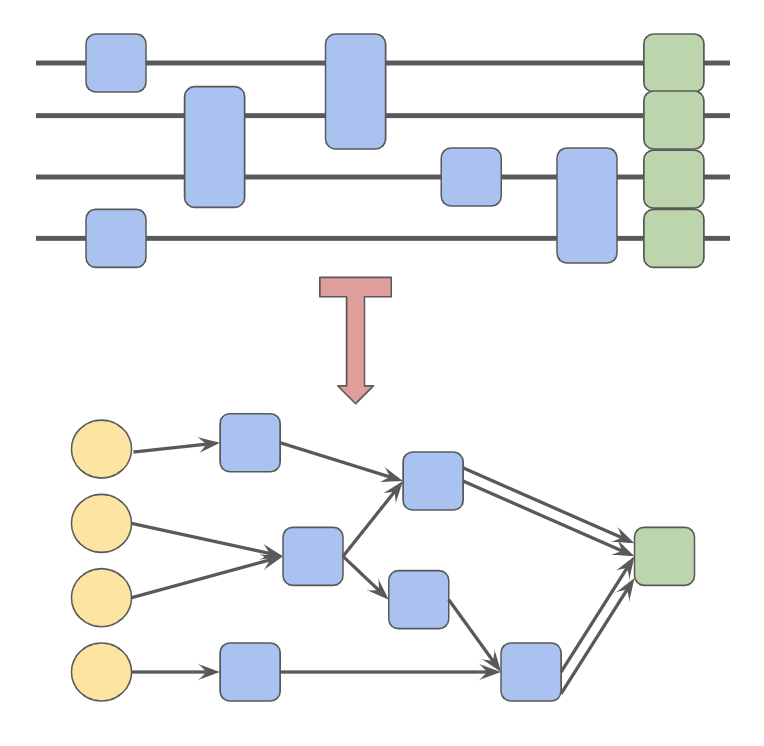}
    \caption{Example mapping from quantum circuit to a directed graph. Each gate is a vertex and an edge connects two vertices if the two gates are connected directly by a qubit wire.}
    \label{fig:mapping}
\end{figure}

Implementing a general circuit cutting routine is a rather tedious task. The \textit{tensor network} formalism becomes convenient for consistent book-keeping \cite{peng2020simulating}. For each quantum circuit, we can map it to a corresponding directed graph: the vertices are quantum gates and an initial set of qubits. Two vertices (gates) are connected if it is directly connected by a qubit wire in the circuit. An example is demonstrated in Figure \ref{fig:mapping}.

The circuit cut then corresponds to removing one edge. And by the graph formalism, we can immediately identify the gates and qubits those gates operate on upstream and downstream of the cut. A subcircuit is produced only when the resulting graph is no longer in one component and this procedure might take multiple cuts (removal of edges) to achieve. For each subcircuit, the associated tensor will be of rank $2k+1$ where $k$ is the number of cuts the subcircuit is involved in. For example, in the two-cut example (Figure \ref{fig_second_case}), subcircuit $A$ can be stored into a rank-3 tensor because it is upstream of one cut. On the other hand, subcircuit $B$ is sandwiched between two cuts, so there are $2*2 + 1 = 5$ indices. To combine two subcircuits, one can use Algorithm \ref{alg:onecut}, \ref{alg:twocut}, or \ref{alg:twocutrand} depending on the situation, so long as the right indices are being contracted.

\section{Experiments}\label{sec:exp}

We give preliminary empirical results for the effectiveness of our proposed approximate circuit cutting scheme. The experiments were run on a personal computer equipped with Intel i5 processors. The algorithms implemented follow exactly as sketched in the pseudocode presented in the previous section. We will show that, despite the simplicity of the program, our method scales well with circuit size both in terms of time complexity and performance. 

To compare two distributions, we propose the following \textit{average variational distance} metric: for distribution $p(x)$ and $q(x)$ over the same sample space $\mathcal X$, the average variational distance is define by the following:
\begin{align}
    \mathcal D(p;q) = \sum_{x \in \mathcal X} \left(p(x) - q(x) \right) \mu(x)
\end{align}
where $\mu(x)$ is the average probability distribution over $\mathcal X$ of $p$ and $q$:
\begin{align}
    \mu(x) = \frac{p(x) + q(x)}{2}
\end{align}
Intuitively, this represents the difference in likelihood functions with respect to the average distribution. This metric effectively captures the goal of approximate circuit cutting: estimating bit strings of large probability. If our estimate distribution has the same shape as the intended distribution, that is, the two distributions share the same high-probability bit strings, then $\mathcal D(p;q)$ will not punish the discrepancy for bit strings in the low likelihood regime. 

\subsection{Case of One Cut}
\begin{figure}
    \centering
    \includegraphics[width=.9\linewidth]{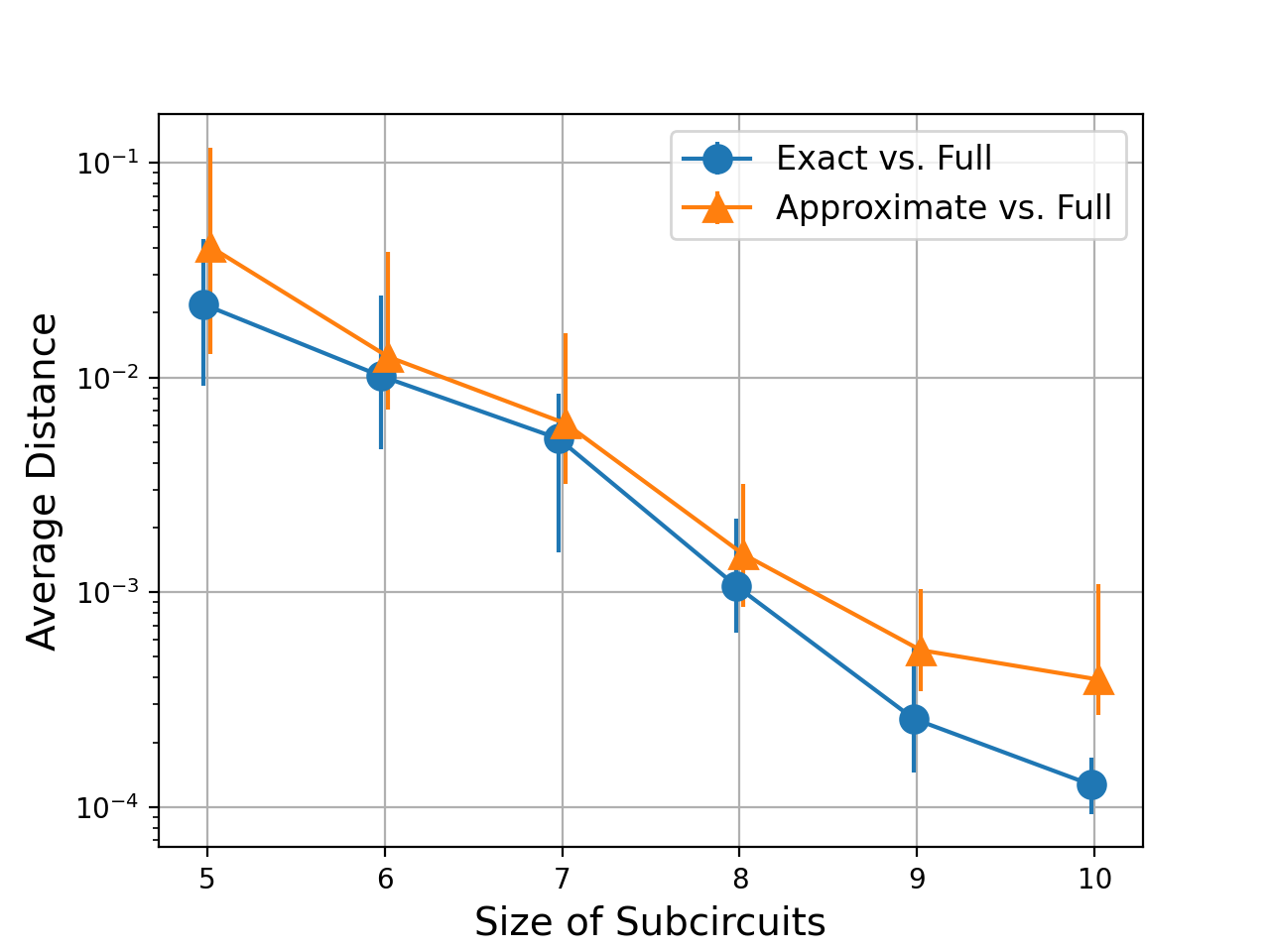}
    \caption{The average distance (in log scale) with respect to the size of each subcircuit in the case of one cut. The error bars show the 25-th and 75-th quantiles.}
    \label{fig:onecut}
\end{figure}

\begin{figure}
    \centering
    \includegraphics[width=.9\linewidth]{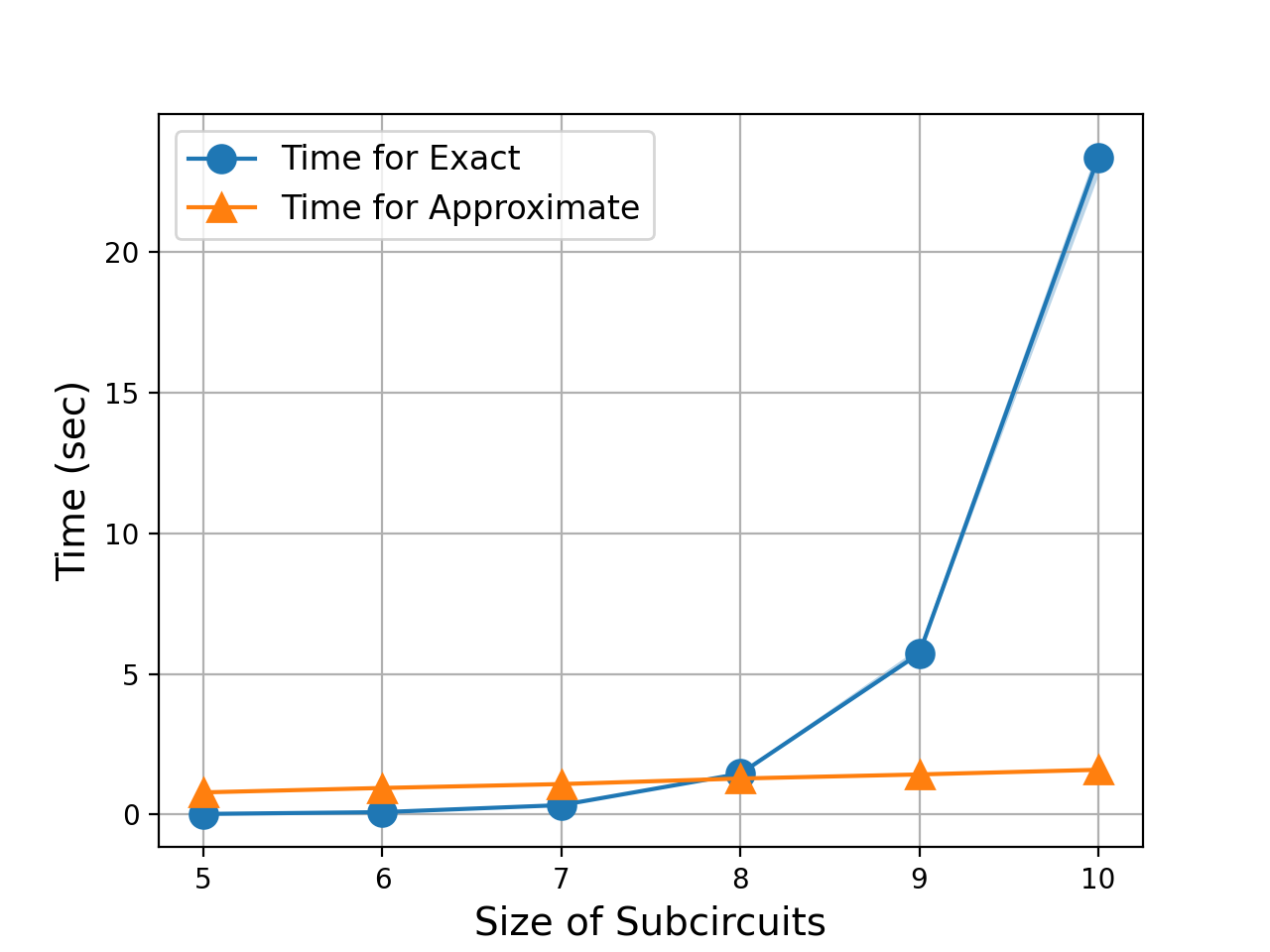}
    \caption{Comparison of running time with respect to the size of subcircuits in the case of one cut.}
    \label{fig:onecuttime}
\end{figure}

For the one-cut case, we generated random subcircuits of the same sizes using Qiskit \cite{cross2018ibm}, then connect the two subcircuits using a CNOT gate. We compared the distributions resulting from the exact reconstruction and the approximate reconstruction against a run of the full circuit without any cuts. This process is repeated for 30 trials, and the results are shown in Figure \ref{fig:onecut}. We can see that the exact reconstruction method consistently outperformed the approximate one, which was expected by virtue of probabilistic methods. However, the discrepancy is not large and overall was close to that of the exact reconstruction. The main advantage of the approximate reconstruction is the computational complexity, which was measured empirically and displayed in Figure \ref{fig:onecuttime}. Here, we linearly increased the number of samples taken with respect to the size of each subcircuit. As a result, the time needed for computation only grows linearly. However, performing exact reconstruction naively will require exponential time. Moreover, the time recorded did not account for time for normalizing the distribution. This shows another advantage of approximate reconstruction: it requires no extra post-processing for normalization. It is important to note that, like all Monte Carlo methods, the more samples are taken, the better the performance is. One could choose to scale the number of samples taken to be more than linear with respect to the size of the circuit if one can afford the time and computational resources.

\subsection{Case of Two Cuts}
\begin{figure}
    \centering
    \includegraphics[width=.9\linewidth]{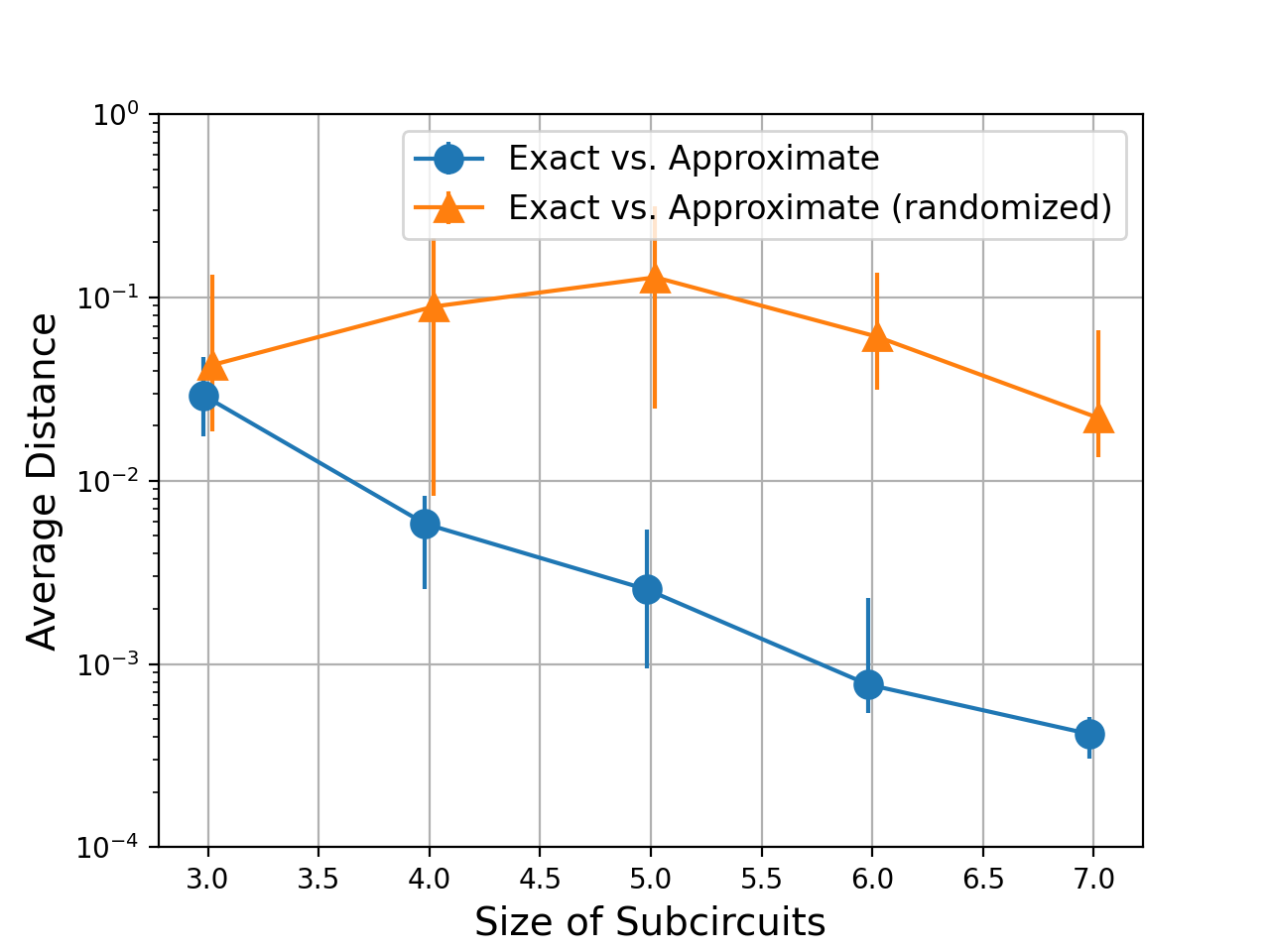}
    \caption{The average distance (in log scale) with respect to the size of each subcircuit for the case of two cuts. The error bars show the 25-th and 75-th quantiles.}
    \label{fig:twocutexp}
\end{figure}

\begin{figure}
    \centering
    \includegraphics[width=.9\linewidth]{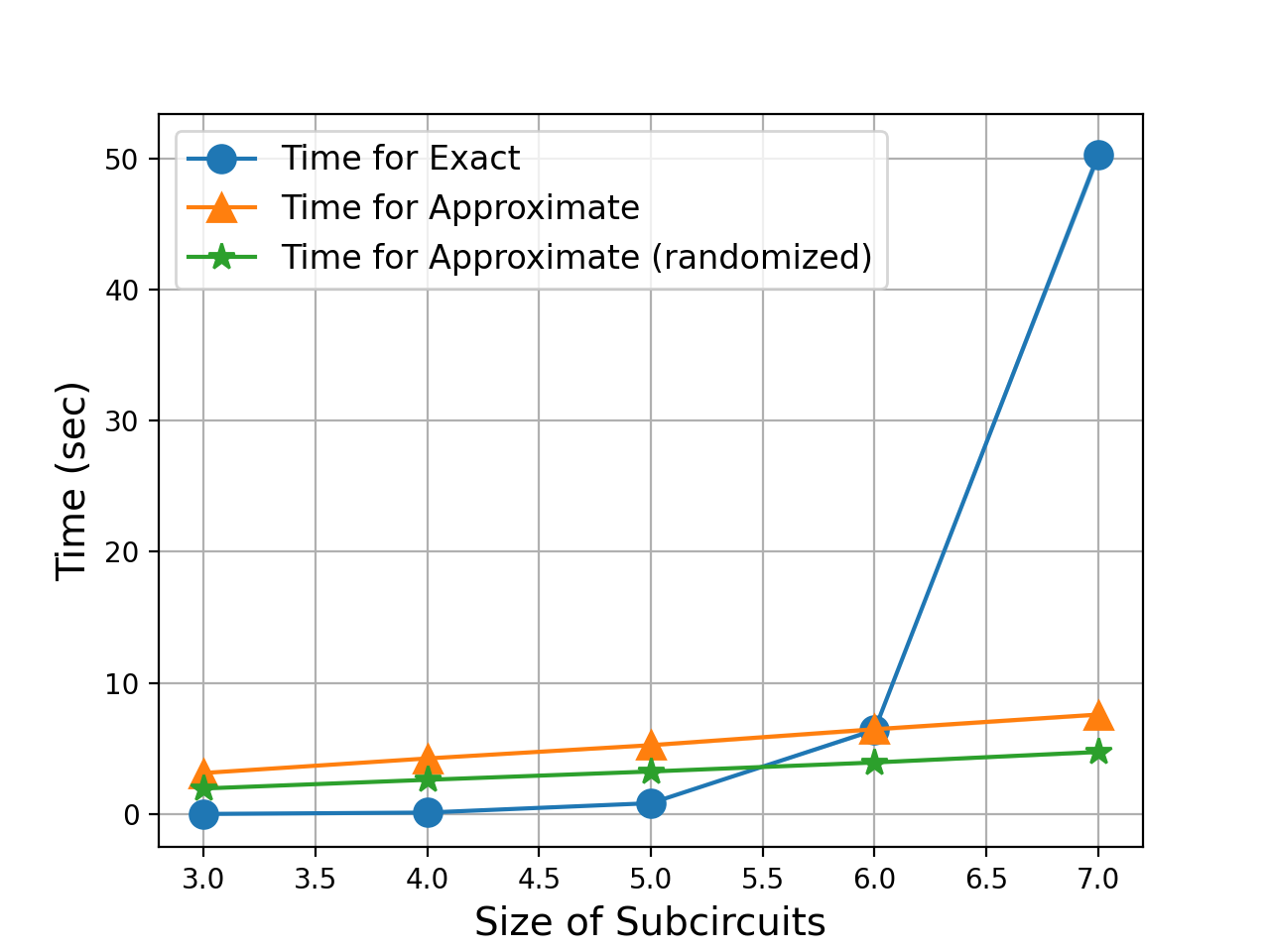}
    \caption{Comparison of running time with respect to the size of subcircuits in the case of two cuts.}
    \label{fig:twocuttime}
\end{figure}

We repeat a similar experiment for the case of two cuts. We generate subcircuits of equal size randomly and connect each subcircuit using a CNOT gate, structured like one depicted in Figure \ref{fig:twocut}. And since circuit cutting generally improves the fidelity \cite{tang2021cutqc}, we will compare the exact reconstruction method with the two proposed approximate methods: one that loops over all indices, and the other randomly selects the index. Again, for each size of subcircuits, the same experiment was repeated 30 times. The results are shown in Figure \ref{fig:twocutexp}. The approximate method without randomized indexing behaved similarly to the case of one cut, showing that the approximate reconstruction procedure can be applied sequentially without significant deterioration in the quality of reconstruction. However, randomly choosing indices resulted in poor accuracy. It is important to keep in mind that randomized indexing was used to escape the exponential scaling in the number of cuts. We suspect that the accuracy of the randomized indexing method can be improved with much larger sample sizes. However, for only a few cuts, sacrificing such magnitude of accuracy is inappropriate. Meanwhile, exponential runtime is again mitigated by controlling the inflation of sample sizes, as in Figure \ref{fig:twocuttime}.

\section{Conclusion and Future Directions}\label{sec:conc}

In this paper, we proposed the concept of \textit{approximate circuit reconstruction}, which aims at piecing together the measurement outcome of subcircuits in a randomized way to avoid unfavorable scaling both in terms of the circuit size and the number of cuts. The experimental results show that this is a promising direction. Below, we list two particular directions to explore. 

\paragraph{Improved Algorithm Design} The current implementation uses the simplest implementation of the Metropolis-Hastings algorithm. Furthermore, no assumptions of quantumness were made throughout. As a result, there is significant room for designing more sophisticated sampling methods that take advantage of existing structures.


\paragraph{Bayesian Reconstruction} Traditionally, MCMC methods are used for Bayesian computation to neglect an intractable normalizing factor that is produced in invoking Bayes' theorem. Here, sampling was used to escape the imperfect normalizing factor from statistical shot noise. Yet, there is potential for further exploiting the properties of MCMC algorithms. For example, integrating priors into reconstructions might be beneficial for reducing noise while keeping computation tractable. However, one would need a probabilistic model for classical distributions generated by quantum circuits, which will also be left as a future direction.

\section*{Acknowledgement}
This work was partially supported by NSF PPoSS under award number 2216923. N.X. was partially supported by the U.S. Army Research Office (ARO)
under award number W911NF1910362.

\bibliographystyle{IEEEtran}
\bibliography{IEEEabrv,ref}

\end{document}